# Unsteady aerodynamic prediction using limited samples based on transfer learning

Wen Ji[1][0000-0002-1733-3835], Xueyuan Sun[1], Chunna Li[1*][ 0000-0003-3476-1505], Xuyi Jia[1][0000-0003-3682-9745], Gang Wang[2][0000-0002-7669-4761] and Chunlin Gong[1][0000-0003-4803-3867]

[1] Shaanxi Aerospace Flight Vehicle Design Key Laboratory, School of Astronautics, Northwestern Polytechnical University, Xi'an, China
[2] School of Aeronautics, Northwestern Polytechnical University, Xi'an, China
chunnali@nwpu.edu.cn

**Abstract.** In this study, a method for predicting unsteady aerodynamic forces under different initial conditions using a limited number of samples based on transfer learning is proposed, aiming to avoid the need for large-scale high-fidelity aerodynamic simulations. First, a large number of training samples are acquired through high-fidelity simulation under the initial condition for the baseline, followed by the establishment of a pre-trained network as the source model using a long short-term memory(LSTM) network. When unsteady aerodynamic forces are predicted under the new initial conditions, a limited number of training samples are collected by high-fidelity simulations. Then, the parameters of the source model are transferred to the new prediction model, which is further fine-tuned and trained with limited samples. The new prediction model can be used to predict the unsteady aerodynamic forces of the entire process under the new initial conditions. The proposed method is validated by predicting the aerodynamic forces of free flight of a high-spinning projectile with a large extension of initial angular velocity and pitch angle. The results indicate that the proposed method can predict unsteady aerodynamic forces under different initial conditions using 1/3 of the sample size of the source model. Compared with direct modeling using the LSTM networks, the proposed method shows improved accuracy and efficiency.

**Keywords:** Unsteady aerodynamic, Limited samples, Transfer learning, Computational fluid dynamics, Long short-term memory network.

## 1   Introduction

Fast and accurate prediction of unsteady aerodynamic forces has various potential applications in the field of aerospace engineering, including aeroelasticity, flight dynamics, and optimization design[1].

In recent years, significant advancements in computational fluid dynamics(CFD), machine learning, and data science have led to the emergence of numerous machine learning-based models for predicting unsteady aerodynamic forces[2]. These models encompass a range of techniques, such as artificial neural network (ANN) models,

Fourier model, LSTM networks [3], and recurrent neural networks(RNN) [4]. Mohammad [5] conducted a comprehensive performance analysis of various models, e.g. ANN model and Fourier model, for predicting unsteady aerodynamic forces. LSTM network is one of the commonly used methods for unsteady aerodynamic modeling. Ayman et al.[6] presented a novel deep learning approach using bidirectional LSTM for accurately predicting unsteady forces associated with wind turbine airfoil pitch motion. Li et al. developed an aerodynamic modeling method based on the LSTM network, capturing the dynamic characteristics of the aerodynamics of an airfoil pitching and plunging in the transonic flow across multiple Mach numbers[7].

Machine-learning-based models for predicting unsteady aerodynamic forces often entail substantial CFD calculations, but the simulation process can be time-consuming and costly. Consequently, several approaches have been proposed to mitigate the computational burden, including transfer learning and data fusion, which aim to reduce the amount of required sample computation. Wang et al. [8] employed a deep transfer learning framework to predict airfoil flow fields with limited data by leveraging airfoil flow fields with sufficient data. Wang et al. proposed a comprehensive framework for dynamic stall prediction by integrating numerical simulations with experimental data, leading to improved results in the unsteady aerodynamic prediction of a NACA0012 airfoil[9]. However, these approaches require large-scale training data, often making data collection the most computationally intensive and time-consuming aspect of the method.

This paper proposes a transfer learning-based method for predicting unsteady aerodynamic forces to reduce the number of required training samples under varying calculation conditions. Initially, a large number of training samples are used to construct a pre-trained network as the source prediction model using an LSTM network. When the initial conditions change, the source prediction model is transferred to the new prediction model, which is further fine-tuned and trained using a limited number of samples to adapt it for prediction under the new initial conditions.

The remaining sections of this paper are organized as follows. Section 2 presents the procedure and mathematical models used in the proposed method. A test case used to verify the proposed method is given in Sect. 3. The summary and conclusions are provided in Sect. 4.

## 2 Proposed Method

This section describes the process of the proposed method, the main methodologies and the models involved in the process.

### 2.1 Process of the proposed method

The modeling and prediction process of the proposed method in this paper is shown in **Fig. 1**. The specific steps are as follows:

1. With the initial conditions provided, high-fidelity simulations are utilized to obtain the source domain data $\mathcal{D}_S = \{x_{t_i}, y_{t_i}\}_{i=1}^{n}$; $x$ and $y$ represent the input and output, respectively, and the input corresponds to the state parameter, while the output corresponds to the aerodynamic force in this work; $t_i (i=1,2,\cdots n)$ is the time point in the simulation;
2. Based on $\mathcal{D}_S$, the initial aerodynamic prediction model $N_{lstm}^{S}$ is built using an LSTM network, and the parameter of the network is $\theta_0$;
3. Changing the initial conditions, high-fidelity simulations with a limited number of time steps are conducted to obtain the target domain $\mathcal{D}_T = \{x_{t_i}, y_{t_i}\}_{i=1}^{m}$, $m < n$;
4. Transfer the initial aerodynamic prediction model $N_{lstm}^{S}$ to the new condition, and it is fine-tuned and retrained using $\mathcal{D}_T$, resulting in the development of a new prediction model $N_{lstm}^{T}$;
5. Use the developed prediction model to predict unsteady aerodynamic forces and moments.

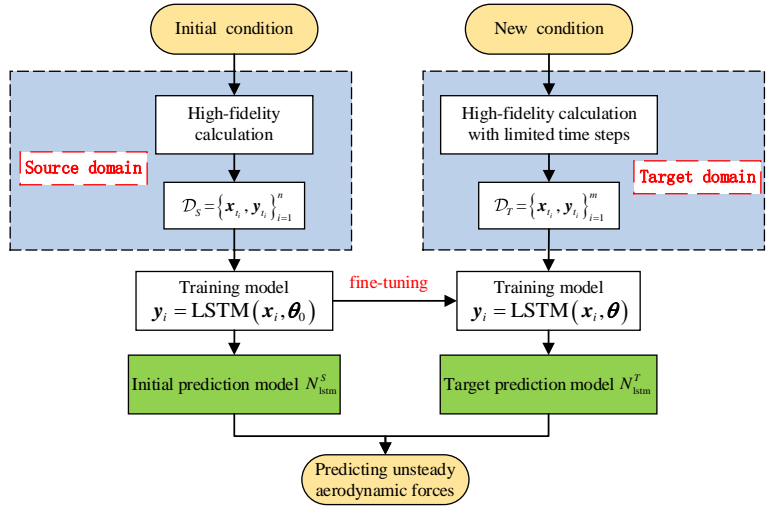

**Fig. 1.** Flow chart of the proposed method

## 2.2  LSTM for unsteady aerodynamic forces prediction

LSTM network is a variant of traditional RNN, which solves the problem of time series classification or regression by learning the long-term correlation between time steps of the sequence data[10]. A typical architecture of an LSTM network is shown in **Fig. 2**.

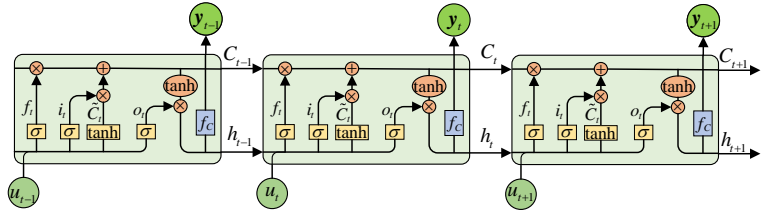

**Fig. 2.** Typical LSTM network architecture

The mathematical model of the LSTM network is provided in the following equations[11],

$$f_t = \sigma\left(W_f \cdot [h_{t-1}, u_t] + b_f\right) \tag{1}$$

$$i_t = \sigma\left(W_i \cdot [h_{t-1}, u_t] + b_i\right) \tag{2}$$

$$\tilde{C}_t = \tanh\left(W_C \cdot [h_{t-1}, u_t] + b_C\right) \tag{3}$$

$$C_t = f_t \otimes C_{t-1} + i_t \otimes \tilde{C}_t \tag{4}$$

$$o_t = \sigma\left(W_o \cdot [h_{t-1}, u_t] + b_o\right) \tag{5}$$

$$h_t = o_t \otimes \tanh(C_t) \tag{6}$$

$$y_t = W_{fc} \otimes h_t + b_{fc} \tag{7}$$

where $f$, $i$, $o$, and $C$ are the forget gate, input gate, candidate gate, and output gate, respectively. Especially, $C$ also represents the cell state, and $\tilde{C}$ represents the updated cell state. $t$ denotes time. $h$ is the short-term memory state. $u$ and $y$ are the input vector and output vector, respectively. $W_f$, $W_i$, $W_C$, $W_o$, and $W_{fc}$ denote the weight matrices. $b_f$, $b_i$, $b_C$, $b_o$, and $b_{fc}$ denote the bias vectors. $\sigma$ and tanh are sigmoid and tanh activation functions.

For unsteady aerodynamic forces prediction, the input variables of the LSTM are the state parameters $S$, such as angle of attack and velocity, whereas the unsteady aerodynamic forces $O$ are considered as the output variables. Thus, the prediction model can be described as

$$O_t = \text{LSTM}\left(S_{t-T+1}, S_{t-T+2}, \ldots, S_{t-1}, S_t\right) \tag{8}$$

where $T$ is the length of the time series.

## 2.3 Transfer learning model

This work uses a pre-training model-based approach to construct a transfer learning model. The pre-training approach is described as follows.

Given the source domain $\mathcal{D}_S = \{x_i, y_i\}_{i=1}^{n_s}$ and the target domain $\mathcal{D}_T = \{x_j, y_j\}_{j=1}^{n_t}$, where $x$ and $y$ represent the input and output, respectively. The number of samples in the source and target domains is denoted as $n_s$ and $n_t$, respectively. Where the source domain is the domain with a large amount of data annotations, and the target domain contains limited data. Pre-training and fine-tuning aim at learning a function $f : x \to y$ parameterized by $\theta$ using the previous parameter $\theta_0$ from historical tasks[12].

For predicting the unsteady aerodynamic forces, the transfer learning model in this work is depicted in **Fig. 3**, illustrating the following process:

1. Based on the source domain $\mathcal{D}_S$, build an initial aerodynamic prediction model $N_{\text{lstm}}^S(\theta_0)$ using the LSTM networks, where the $\theta_0$ is the parameter of the network;

2. Changing the calculation conditions, the target domain $\mathcal{D}_T = \{x_{t_i}, y_{t_i}\}_{i=1}^m$ is obtained by $m$ time steps high-fidelity simulations;

3. Transfer the model $N_{\text{lstm}}^S(\theta_0)$ to the target domain, then fine-tune the layer parameters and retrain the network through $\mathcal{D}_T$;

4. Following steps 1-3, the new prediction model $N_{\text{lstm}}^T(\theta)$ in the target domain is generated, and the $\theta$ is the parameter of the new network.

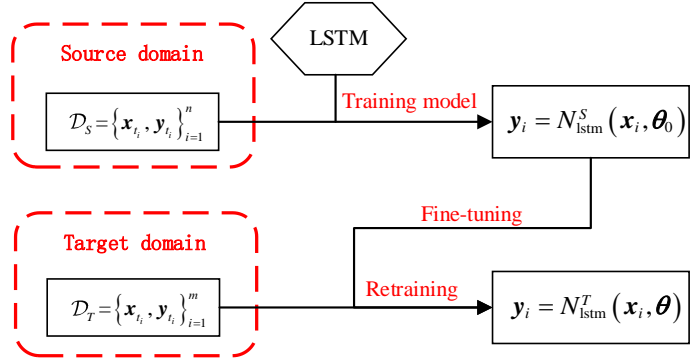

**Fig. 3.** Transfer learning model based on the pre-training method

## 3 Test case

### 3.1 Calculation Model

An ARL spinning projectile configuration is selected as the computational model to validate the proposed aerodynamic prediction method. **Fig. 4** shows the geometry, mass, and rotational inertia parameters of the projectile, and cg represents the center of mass.

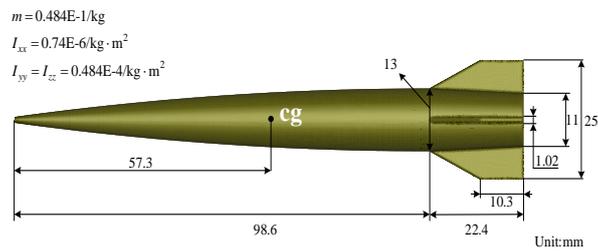

**Fig. 4.** Spinning Projectile Configuration

To accurately calculate the unsteady aerodynamic forces and the flight trajectory of the projectile[13], coupled CFD/RBD simulation is used in this work. Detailed information regarding the coupled CFD/RBD method and accuracy verification of the in-house flow solver can be found in Ref.[14].

The computational grid is an unstructured hybrid grid, as shown in **Fig. 5**. The boundary layer grid employs triangular prism elements, with the first layer height being $1\times10^{-6}$ m, a total of 25 layers, and a growth rate of 1.2. Tetrahedron elements are used in other areas. The entire set of grids comprises 20,816 surface grids and 811,923 volume grids. The turbulence model is Sparlat-Allmaras one-equation model.

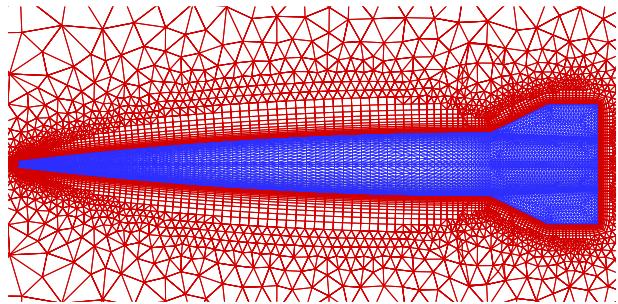

**Fig. 5.** Hybrid unstructured mesh near the projectile

The initial state of the trajectory simulation is shown in Table 1. The equivalent free-flight Mach number is 3; the Reynolds number is $7.083\times10^7$, and the angle of attack is 4.9 deg.

Table 1. Initial condition of the spinning projectile

| Direction | Position/m | Velocity/(m/s) | Attitude angle/rad | Angular rate/(rad/s) |
|---|---|---|---|---|
| X | 4.593 | 1030.81 | 2.051 | 2518.39 |
| Y | -0.2 | 22.064 | 0.088 | 52.802 |
| Z | -0.159 | 86.278 | -0.023 | 22.233 |

### 3.2 Initial prediction model establishment

The CFD/RBD coupling simulations were performed based on the initial conditions in Table 1, with a total of 2500 time steps.

The input and output of the prediction model in Eq.8 can be written as follows.

$$\begin{aligned} \mathbf{S} &= [u,v,w,\psi,\varphi,\theta,p,q,r,\dot{\psi},\dot{\varphi},\dot{\theta}] \\ \mathbf{O} &= [F_x,F_y,F_z,M_x,M_y,M_z] \end{aligned} \quad (9)$$

where $F$ is the force; $M$ represents the moment. $u$, $v$, and $w$ represent the respective components of the velocity vector in the body axis system, while $p$, $q$, and $r$ denote the components of the angular velocity vector in the body reference frame. $\varphi$, $\psi$ and $\theta$ represent roll, yaw, and pitch angle, respectively. Here, the length of the time series $T$ = 10.

The first 85% of the data is the training set and the last 15% is the test set with no validation set to train and obtain the initial prediction model, and the network is trained with Nvidia GeForce RTX 3080.

The root mean square error (*RMSE*) and mean absolute error (*MAE*) are used to evaluate the prediction performance of the model, defined as follows. $N$ denotes the number of samples; $y_i$ represents the real value obtained by CFD, whereas $\hat{y}_i$ is the value predicted by the prediction model.

$$\begin{aligned} RMSE &= \sqrt{\frac{1}{N}\sum_{i=1}^{N}(y_i-\hat{y}_i)^2} \\ MAE &= \frac{1}{N}\sum_{i=1}^{N}|y_i-\hat{y}| \end{aligned} \quad (10)$$

### 3.3 Prediction in different initial conditions

With all other conditions held constant, change the initial angular velocity in the x-direction in Table 1 to 2000 rad/s and the initial pitch angle to 2 deg. Based on the new initial condition, CFD/RBD coupled simulations with 2500-time steps are performed. The LSTM network of the initial prediction model was transferred to this state, and the network was tuned and trained using the data from the first 25%-time steps. The tuned network is trained for a total of 1000 steps to obtain a new prediction model. For comparison, the first 25% of the data was used as a training set to directly build a prediction model using the LSTM network.

**Fig. 6** illustrates a comparison of the predicted values obtained directly using the LSTM network, the predicted values using the proposed method, and the real values obtained from the CFD/RBD coupled simulation. The horizontal coordinate $x_g$ represents the displacement of the projectile's center of mass along the trajectory in the direction of the x-axis in the inertial coordinate system. The error comparison of different methods and the time required for network training are shown in **Table 2**.

Based on **Fig. 6**, it is evident that, with limited samples, the predicted values obtained using the proposed method are closer to the real values compared to the direct use of LSTM networks. With the exception of the axial force $F_x$ and the rolling moment $M_x$, predicted values of all other forces and moments using the proposed method are close to the real values, and the accuracy of the predicted values for $F_x$ and $M_x$ has significantly improved compared to directly using the LSTM networks.

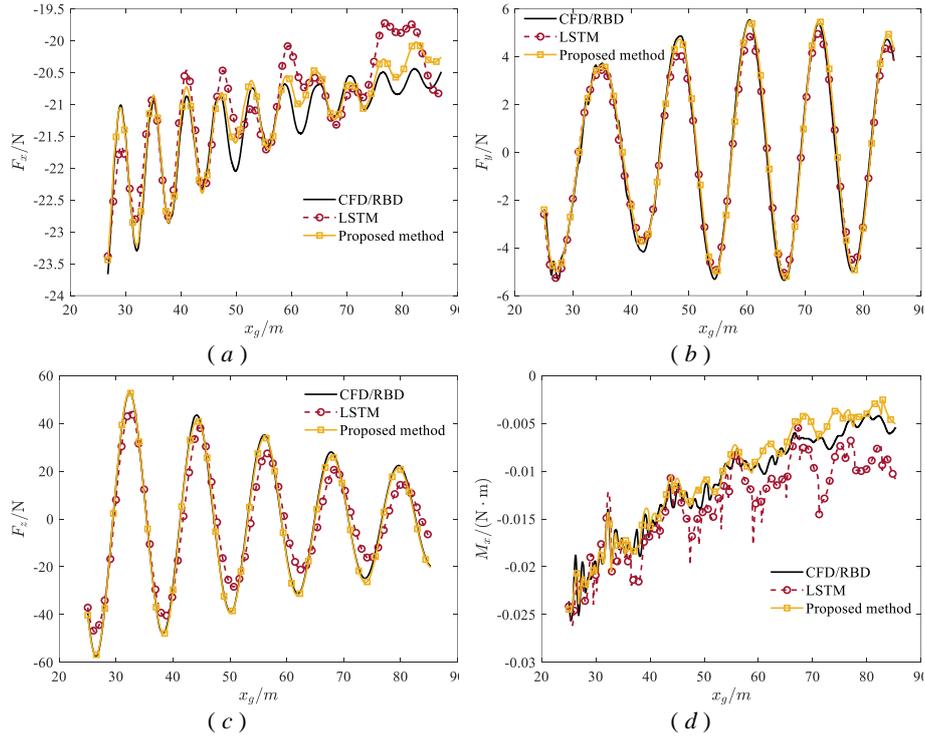

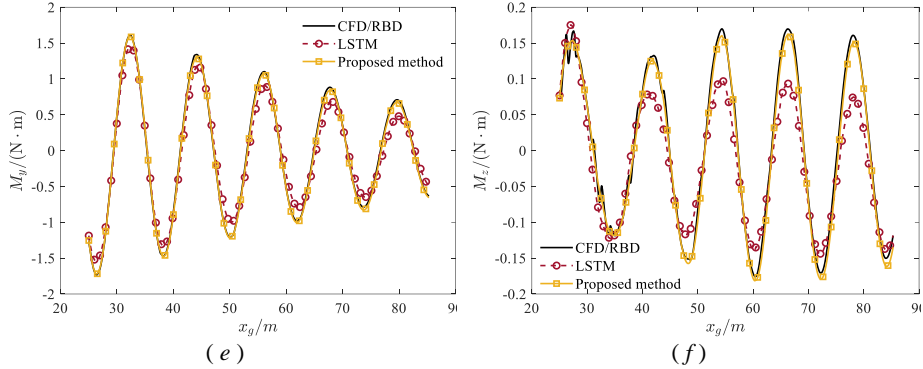

**Fig. 6.** Comparison of predicted and true values of different methods

From **Table 2**, the time consumption for fine-tuning the network of the initial prediction model and training it using the proposed method is decreased by 41% compared to directly using the LSTM networks. Compared to the direct modeling using LSTM networks, the proposed method shows reduced *MAE* and *RMSE* in the predicted values, particularly for the lateral force $F_z$, with reductions of 86.59% and 99.72%, respectively, indicating a significant improvement in prediction accuracy. Based on the above analysis, it is evident that the proposed method can maintain good prediction accuracy when the initial conditions are changed and the training samples are limited.

**Table 2.** Comparison of prediction error and training time consumption of different methods

| Method | Time-consumption | Error | $F_x$ | $F_y$ | $F_z$ | $M_x$ | $M_y$ | $M_z$ |
|---|---|---|---|---|---|---|---|---|
| Proposed method | 25min18s | *RMSE* | 0.005 | 0.007 | 0.024 | 2.88e-4 | 6.25e-4 | 2.20e-4 |
| | | *MAE* | 0.168 | 0.251 | 0.888 | 0.001 | 0.024 | 0.008 |
| LSTM | 42min5s | *RMSE* | 0.011 | 0.009 | 0.179 | 6.18e-4 | 0.004 | 8.89e-4 |
| | | *MAE* | 0.361 | 0.332 | 7.214 | 0.002 | 0.136 | 0.032 |

## 4   Conclusions

This study develops a method for predicting unsteady aerodynamic forces using limited samples based on the LSTM network and transfer learning, achieving aerodynamic forces prediction with limited samples under new initial conditions by transferring the network of the source model to the new prediction model. The accuracy and efficiency of the proposed method are validated through the prediction of aerodynamic forces in the free flight of a high-spinning projectile considering the large variation of the initial conditions. Conclusions can be drawn as follows:

(1) Once the source model has been established, the proposed method can predict unsteady aerodynamic forces under different initial conditions using less than 1/3 of the sample size of the source model;
(2) With the same amount of training samples, the modeling time of the proposed method is reduced by 41% compared to the direct modeling using LSTM networks, and the prediction errors are also decreased.

Moreover, the proposed method still requires a small number of high-fidelity simulations to obtain the samples, and the selection of the samples will directly affect the prediction accuracy. Thus, the adaptive adjustment of the network parameters according to the target-domain data will be one of the focuses of the future.

**Acknowledgments** This work was supported by the National Natural Science Foundation of China (No. U2141254).